\documentclass[aps,prl,twocolumn,showpacs,showkeys,preprintnumbers,amsmath,amssymb,superscriptaddress,floatfix]{revtex4}

\usepackage{graphicx}
\usepackage{dcolumn}
\usepackage{bm}
\usepackage{hyperref}

\begin{document}

\newcommand{\sq}{\xi}
\newcommand{\sqc}{\xi}

\title{Enhanced and reduced atom number fluctuations in a BEC splitter}

\author{Kenneth Maussang}\email{Kenneth.Maussang@lkb.ens.fr}
\affiliation{Laboratoire Kastler-Brossel, ENS, Universit\'e Pierre et Marie Curie-Paris 6, CNRS, 24 rue Lhomond, 75\,231 Paris Cedex 05, France}
\author{G. Edward Marti}\altaffiliation[Present address: ]{Department of Physics, University of California, Berkeley, California 94720, USA.}
\affiliation{Laboratoire Kastler-Brossel, ENS, Universit\'e Pierre et Marie Curie-Paris 6, CNRS, 24 rue Lhomond, 75\,231 Paris Cedex 05, France}
\author{Tobias Schneider}\altaffiliation[Present address: ]{Institut f\"ur Experimentalphysik, Heinrich Heine Universit\"at, 40225 D\"usseldorf, Germany.}
\affiliation{Laboratoire Kastler-Brossel, ENS, Universit\'e Pierre et Marie Curie-Paris 6, CNRS, 24 rue Lhomond, 75\,231 Paris Cedex 05, France}
\author{Philipp Treutlein}
\affiliation{Departement Physik, Universit\"at Basel, Klingelbergstrasse 82, CH-4056 Basel, Switzerland}
 \author{Yun Li}
\affiliation{Laboratoire Kastler-Brossel, ENS, Universit\'e Pierre et Marie Curie-Paris 6, CNRS, 24 rue Lhomond, 75\,231 Paris Cedex 05, France}
\affiliation{Precision Spectroscopy, Department of Physics, East China Normal University, Shanghai 200062, China}
\author{Alice Sinatra}
\affiliation{Laboratoire Kastler-Brossel, ENS, Universit\'e Pierre et Marie Curie-Paris 6, CNRS, 24 rue Lhomond, 75\,231 Paris Cedex 05, France}
\author{Romain Long}
\affiliation{Laboratoire Kastler-Brossel, ENS, Universit\'e Pierre et Marie Curie-Paris 6, CNRS, 24 rue Lhomond, 75\,231 Paris Cedex 05, France}
\author{J\'er\^ome Est\`eve}
\affiliation{Laboratoire Kastler-Brossel, ENS, Universit\'e Pierre et Marie Curie-Paris 6, CNRS, 24 rue Lhomond, 75\,231 Paris Cedex 05, France}
\author{Jakob Reichel}
\affiliation{Laboratoire Kastler-Brossel, ENS, Universit\'e Pierre et Marie Curie-Paris 6, CNRS, 24 rue Lhomond, 75\,231 Paris Cedex 05, France}

\date{\today}
\begin{abstract}
  We measure atom number statistics after splitting a gas of ultracold
  $^{87}$Rb atoms in a purely magnetic double-well potential created
  on an atom chip. Well below the critical temperature for
  Bose-Einstein condensation $T_c$, we observe reduced fluctuations
  down to {$-4.9$}\,dB below the atom shot noise level. Fluctuations
  rise to more than $+3.8\,$dB close to $T_c$, before reaching the
  shot noise level for higher temperatures. We use two-mode and
  classical field simulations to model these results. This allows us
  to confirm that the super-shot noise fluctuations directly originate
  from quantum statistics.
\end{abstract}
\maketitle

Since the achievement of Bose-Einstein condensation (BEC) in dilute
atomic gases, different experimental techniques have been developed in
order to coherently split a BEC into two spatially separate
parts~\cite{Ketterle,Anderson,Schumm,Oberthaler_Gati}, with atom
interferometry as one of the motivations. Even though BECs are usually
in the weakly interacting regime, the interactions between the
particles dramatically affect the physics of the splitting. In
particular, repulsive interactions limit the phase coherence between
the two split parts~\cite{Phase}, but also reduce atom number
difference fluctuations, giving rise to non-classical squeezed
states~\cite{Kitagawa,Squeezing,Squeezing_Ketterle,BPhillipps,Squeezing_MKO}.

\begin{figure}[t]
\includegraphics[scale=1]{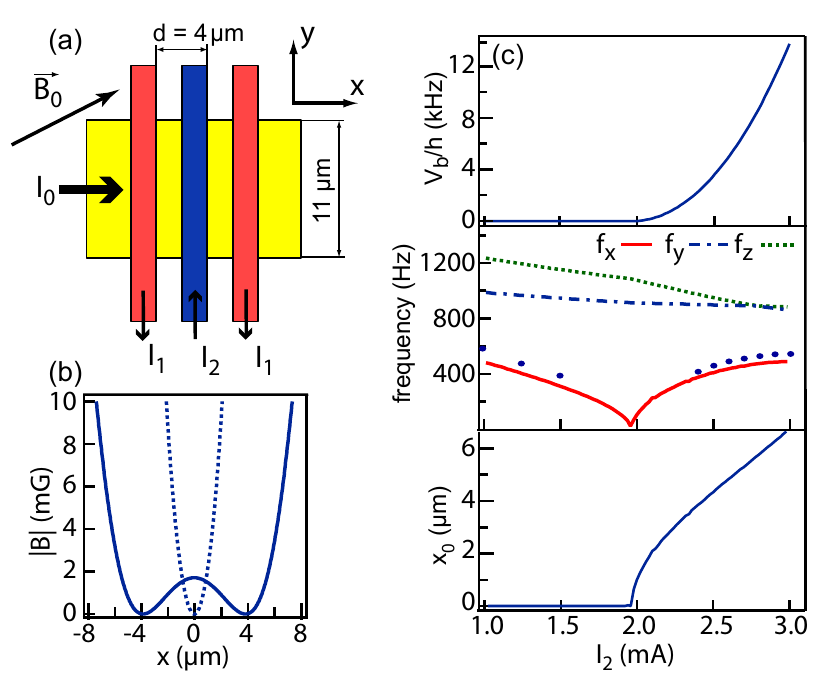}
\caption{\label{fig:ChipPotential} (color online). (a) Chip geometry:
  we use DC currents superposed with a static homogeneous field
  $\mathbf{B_0}=(B_x,B_y,B_z)=(15,8,0)$\,G.  To first approximation,
  the potential can be understood as follows: The field of
  $I_0=100\,$mA combines with $\mathbf{B_0}$ to create harmonic
  confinement in the $xy$ plane. The minimum is located at a distance
  $z_0=\mu_0 I_0/2\pi B_y$ from the $I_0$ wire and the field in the
  minimum is $B_x$. This minimum is modified by $I_1=2.0$\,mA and
  $I_2$, whose fields are predominantly along $x$ on the trap
  axis. For $z_0\gtrsim d$, the two $I_1$ wires together provide
  harmonic confinement along $x$. $I_2$, of opposite polarity, creates
  the barrier of adjustable height and also determines the spacing
  between the two resulting wells.  (b) Profile of the trapping
  potential along the splitting axis $x$, for $I_2=0$\,mA (cooling
  trap, dashed line) and $I_2=2.4$\,mA (solid line). The $x$ axis is
  the same as in (a). (c) Barrier height $V_b$ (top), trap frequencies
  $(f_x,f_y,f_z)$ (center) and position $x_0$ of the {right} minimum
  (bottom) as functions of the current $I_2$. Lines: numerical
  calculation based on current configuration. Blue circles: measured
  frequencies $f_x$.}
\end{figure}

In this Letter, we use a purely static magnetic potential created on
an atom chip to realize a nonlinear spatial ``beam'' splitter for a
BEC. We investigate the physics of the splitting and focus on atom
number fluctuations and the role of temperature. At low temperatures,
where the interaction energy dominates, we directly observe number
squeezed states with relative population fluctuations $-4.9$\,dB below
shot noise, as first shown in~\cite{Squeezing_MKO} and indirectly
observed in~\cite{Squeezing_Ketterle}. The two separated but weakly
linked parts of the BEC constitute a bosonic Josephson junction,
usually described by a two mode model (TMM)~\cite{TMM_Ananikian}. Our
results are in agreement with the TMM, which also predicts that the
observed squeezing is accompanied by high phase coherence.

The magnetic trap configuration allows barrier heights up to several
$\mu$K and straightforward evaporative cooling, so that we can
separate clouds with increasing temperature all the way to the
non-degenerate regime. In the intermediate temperature regime, where
both a significant condensate and thermal fraction are present, we
observe large super-binomial fluctuations in the number difference
between the two parts. This excess of fluctuations is a direct
signature of the Bose statistics, in close analogy to the bunching
effect in quantum optics \cite{EsteveBouchoule_bunching}. 

Close to the BEC transition, the condensates show significant
depletion and the TMM breaks down.
We complement our experiments by a theoretical investigation of this
regime using a classical field approach and show that large
super-binomial fluctuations are a general feature at thermal
equilibrium. Although the experiments are not performed at
equilibrium, our observations are still in qualitative agreement
with these theoretical results.

Our experiment uses a two-layer atom chip to prepare a $^{87}$Rb BEC
in the $\left|\mbox{F}=2,m_{\text{F}}= 2\right\rangle$ hyperfine
state, and then split it in a double-well potential with adjustable
barrier height and well spacing.  The magnetic potential along the
weak axis is well approximated by $V(x)=V_b[1-(x/x_0)^2]^2$, where
$V_b$ is the barrier height and $2\,x_0$ is the distance between the
wells (Fig.~\ref{fig:ChipPotential}, (a) and (b)). The transverse
potential is harmonic with trap frequencies $\omega_{y,z}/2\pi \approx
1\,\textrm{kHz}$. Unlike RF dressed potentials, which tend to create
very elongated traps \cite{Squeezing_Ketterle,Schumm}, our traps have
an aspect ratio close to 1, which strongly suppresses phase
fluctuations within each well. We split the trap by increasing $I_2$,
which simultaneously increases both $V_b$ and $x_0$
(Fig.~\ref{fig:ChipPotential}(b)-(e)).  The interesting regime of two
weakly coupled condensates occurs for a barrier height on the order of
the chemical potential (few kHz), corresponding to a current
$I_2\simeq2.5$\,mA in our experiments. In this region, the tunneling
rate is divided by $2$ when the barrier $V_b/h$ increases by
$75$\,Hz. We estimate fluctuations of the barrier height to be on the
order of $20$\,Hz. A magnetic field component normal to the chip leads
to a small energy difference $\Delta E$ between the two potential
minima. A magnetic shield with an attenuation factor of $\sim 30$
reduces fluctuations of $\Delta E$ to a few Hz.

\begin{figure}[t]
\begin{center}
\includegraphics[scale=1]{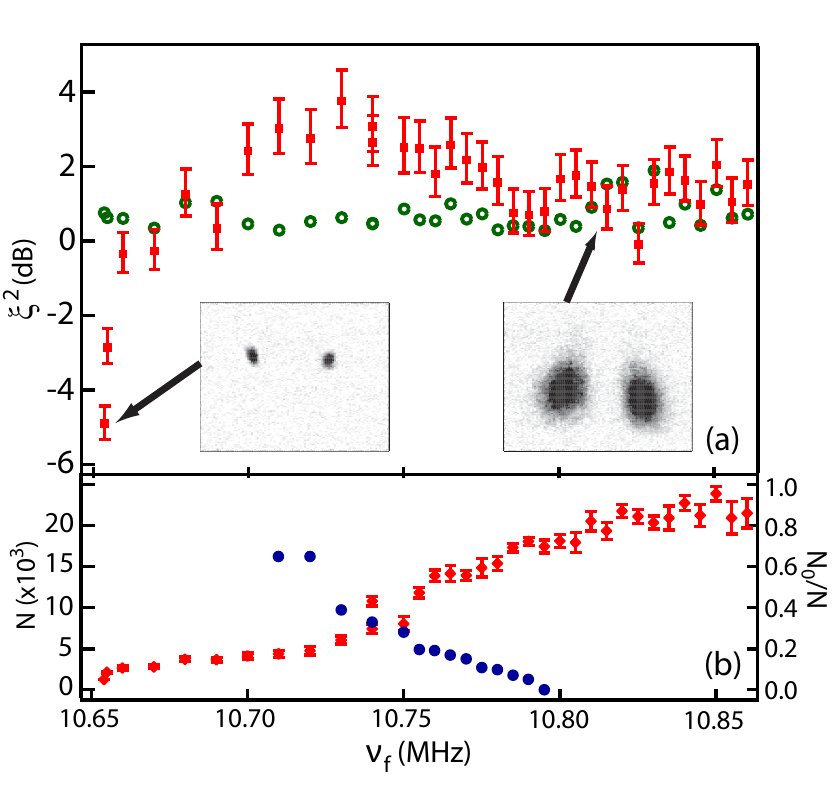}
\end{center}
\caption{\label{fig:FluctRF} (color online). Second experiment: Effect
  of the temperature on $\sqc^2$. We vary the final frequency $\nu_f$
  of the evaporative cooling ramp before splitting the cloud using a
  $50$\,ms linear current ramp. Typical absorbtion images are shown in
  the inset of (a). Red squares: $\sqc^2$. Green circles: fringe
  noise. (Estimation of the $\sqc^2$ we expect to measure when the
  actual fluctuations are binomial - the estimate is obtained by
  analyzing a region of the image that does not contain atoms.)  (b)
  Total atom number $N$ (red diamonds) and condensed fraction $N_0/N$
  prior to splitting (blue circles).}
\end{figure}

The expected atom number fluctuations are on the order of a few tens
of atoms, placing stringent requirements on the imaging system. We
perform absorption imaging using a back-illuminated CCD camera and
3\,ms delay between absorption and reference images. The total quantum
efficiency including optical losses is $q=0.84$. Probe pulses have
$\tau=50\,\mu$s duration and an intensity close to saturation. Atom
numbers are calibrated following a procedure inspired
by~\cite{DGO}. We estimate our total systematic error to be at most
$18$\%. Atoms in the two wells are resolved by applying a large
gradient with the central conductor ($I_2$) for the first
$50\,\mu\mathrm{s}$ of a $6\,\mathrm{ms}$ time-of-flight.  Photon shot
noise, scaling as the root of the image area, leads to a standard
deviation of about 20 atoms for a BEC of $10^3$ atoms imaged over
$800$ pixels, and about 60 atoms for a thermal cloud ($2\times10^4$
atoms, $7000$ pixels), pixel area being $9.5\,\mu$m$^2$ in the object
plane. This is close to the theoretical limit $\sqrt{16/q \sigma
  \Gamma \tau} = 0.19\,\mu\mathrm{m}^{-1}$, where $\sigma$ is the
scattering cross section and $\Gamma$ the natural linewidth, and
always remains below atom shot noise; furthermore, this noise is
well-characterized and stable so that it can be subtracted.  Technical
fluctuations (moving fringes) constitute a second noise source, which
is negligible for small BECs, but becomes comparable to photon shot
noise for large atomic clouds. We do not attempt to remove this noise,
but indicate an estimate of its level (green dots in
Fig.~\ref{fig:FluctRF}).

To measure the fluctuations of the atom number difference $N_L-N_R$
between the left and the right well, we acquire a dataset consisting
of a large number of absorption images (typically $>100$) taken under
identical conditions. In each image, we measure $N_L$ and $N_R$ (see
inset of Fig.~\ref{fig:FluctRF}). To compensate for a slight imbalance
in the splitting, we determine the probability to be in the left
(right) well $f_{L,R}=\langle N_{L,R}/N\rangle$, where $N=N_L+N_R$ and
the average is taken over all images. For each image, we calculate
$n=f_RN_L-f_LN_R$, which is the deviation of $(N_L-N_R)/2$ from the
expected value $(f_L-f_R)(N_L+N_R)/2$. We define the number squeezing
factor as the variance of $z=n/\sqrt{f_Lf_R(N_L+N_R)}$. Correcting for
the photon shot noise contribution $z_p$ \footnote{$z_{p} =
  \sqrt{({f_R^2 (\delta N_L)^2 + f_L^2 (\delta N_R)^2})/({f_L f_R (N_L
      + N_R)})}$, where $\delta N_L, \delta N_R$ are the atom number
  uncertainties in each well due to photon shot noise.} leads to the
final expression $\sq^2 = \langle z^2 \rangle-\langle z_p^2 \rangle$.
This definition is first-order insensitive to fluctuations in the
total number of atoms and produces $\sq^2 = 1$ for a binomial
distribution.
In our data, the correction $\langle z_p^2\rangle$ is always smaller
than $\langle z^2\rangle$ itself.

\begin{figure}[t]
\includegraphics[scale=1]{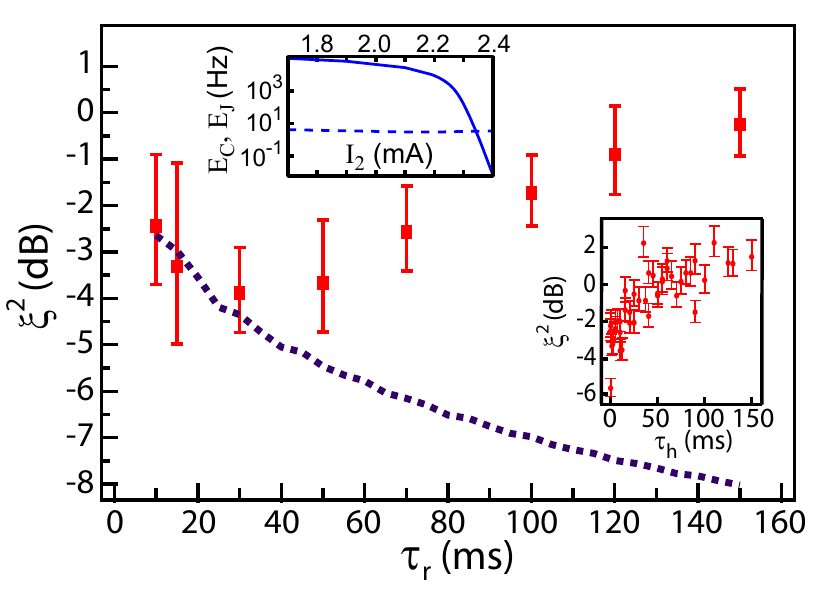}
\caption{\label{fig:FluctSpeed} (color online).  First experiment:
  Fluctuations of the atom number difference as a function of
  splitting time $\tau_r$ for a BEC with N=1300. Red squares: measured
  $\sqc^2$; dotted line: dynamical TMM simulation.  Upper inset: $E_C$
  (dashed line) and $E_J$ (solid line) from numerical 3D GPE solution
  for 1300 atoms using the full calculated potential. Lower inset:
  Degradation of the number squeezing under the influence of
  heating. The BEC is held for a variable time $\tau_h$ before
  splitting; $\tau_r=50$\,ms is constant.}
\end{figure}

In a first experiment, we split an almost pure BEC of $1300$ atoms and
investigate the influence of the time $\tau_r$ during which the
barrier is raised by increasing $I_2$ from zero to 3.9\,mA, well above
the chemical potential. The final number squeezing $\sqc^2$ is shown
in Fig.~\ref{fig:FluctSpeed}. We observe a decrease of fluctuations
below the shot noise limit with increasing $\tau_r$ up to $50\,$ms,
followed by a slow increase for longer times.  This agrees with the
expectation that an optimum should exist between very short $\tau_r$
creating excitations in the BEC and/or not leaving sufficient time for
tunnelling, and very long $\tau_r$ where heating and atom loss become
important. For a quantitative analysis, we start by noting that
low-energy excitations from the symmetric many-particle ground state
correspond to populating the first excited mode, which is the
longitudinal dipole mode for $V_b=0$ and becomes the Josephson plasmon
mode as the barrier raises.  These excitations are spaced by $\hbar
\omega_p$, where the plasma frequency $\omega_p$ decreases from
$\omega_p=\omega_x$ to 0 as the barrier raises \cite{Squeezing}.  The
interesting dynamics can be expected to occur when the barrier height
approaches the chemical potential. In this region, $\hbar\omega_p$ is
already much smaller than $\hbar\omega_x$, which is the energy scale
of the next higher modes. Therefore, a TMM is expected to work well in
this region at least as long as $k_B T < \hbar\omega_x$. The two
parameters entering the TMM are the charging energy $E_C$ accounting
for the interaction between the particles and the Josephson energy
$E_J$ that characterizes the tunnelling. We calculate them for each
barrier height by solving the 3D Gross-Piteavskii equation for the
experimental trap \cite{TMM_Ananikian}.  In the TMM simulations, we
describe the initial state before splitting by a thermal density
matrix, which we evolve according to the von Neumann equation.  We
start the simulation at $I_2=1.9\,$mA, slightly below the splitting
point.  We have checked that the results depend only weakly on the
starting point (initial $I_2$) in this region. An initial temperature
$T_i=50$\,nK reproduces the measured fluctuations after a ramp of
$10$\,ms. The $T_i$ thus found is then used for all longer $\tau_r$.
The simulation reproduces well the observed squeezing for $\tau_r$ up
to 30\,ms.  For $\tau_r>40$\,ms, the experimental data shows a
degradation of squeezing which we attribute to technical heating. To
confirm this hypothesis, we hold the BEC for a variable time $\tau_h$
before splitting it with a $50$\,ms ramp (inset of
Fig.~\ref{fig:FluctSpeed}).

In a second experiment, we investigate the effect of the temperature
on the squeezing. We ramp $I_2$ linearly from 0 to $3.9$\,mA in
50\,ms, and vary the temperature by changing the final frequency
$\nu_{f}$ of the evaporative cooling ramp. The results are plotted in
Fig.~\ref{fig:FluctRF}. Below 10.73\,MHz, i.e. well below $T_c$, we
observe a crossover from sub- to super-binomial fluctuations with
increasing temperature. Around $10.73$\,MHz, the condensed and thermal
fractions are comparable, and we observe large super-poissonian
fluctuations with a maximum of $+3.8$\,dB. As the temperature
increases further, fluctuations decrease, and level when the
condensate fraction reaches zero. The measured asymptotic level of
$1$\,dB is consistent with binomial statistics given our fringe noise
(Fig.~\ref{fig:FluctRF}(a), green circles).  According to the TMM
calculation, our best result of $\sqc^2=-4.9\,^{+0.5}_{-0.4}$\,dB
corresponds to a phase coherence $\langle \cos \varphi \rangle
\sim0.93$ immediately after the splitting, where $\varphi$ is the
relative phase between the two clouds. This would result in a possible
metrology gain of $-4.4$\,dB compared to the standard quantum limit
using this state in an atom interferometer.

\begin{figure}[t]
\begin{center}
\includegraphics[scale=1]{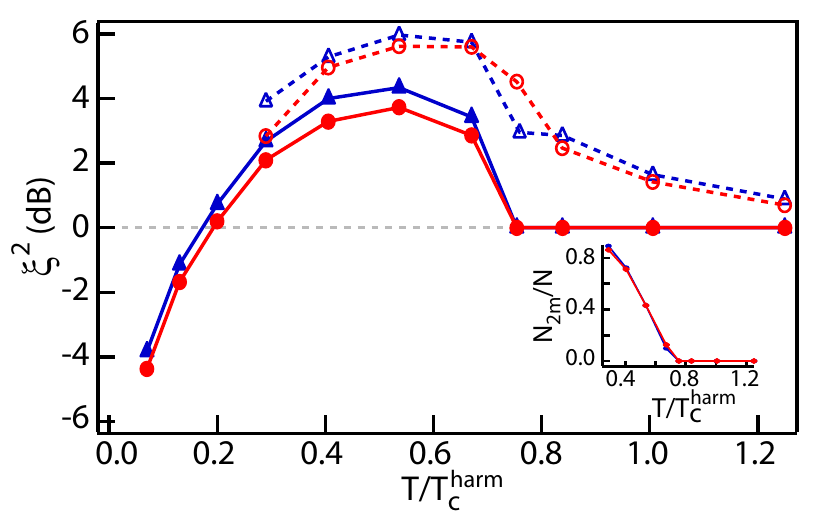}
\end{center}
\caption{\label{fig:theo} (color online). Classical field simulations
  (open symbols+dotted lines) and modified TMM (full
  symbols+lines). Blue {lines and} triangles: $N=6000$; red {lines
    and} circles: $N=17000$. The trapping potential is harmonic
  transversally ($f_{y,z}=1120,1473$\,Hz), and quartic along $x$ with
  $V_b/h=2.69$\,kHz and $x_0=3.8$\,$\mu$m. The longitudinal
  oscillation frequency within each well is $f_x=411$\,Hz. The
  temperature unit in the figure is the ideal gas transition
  temperature in the harmonic potential prior to splitting $T_c^{\rm
    harm}$ ($N=6000$: $T_c^{\rm harm}= 0.89\,\mu$K, $N=17000$:
  $T_c^{\rm harm}=1.25\,\mu$K). Inset: Fraction of the population in
  the two lowest energy modes $N_{2m}$, extracted from the classical
  field simulations.}
\end{figure}

The observed sub- and super-binomial regimes originate from the
interplay between interactions and quantum statistics. Lowering the
temperature, the onset of super-binomial fluctuations occurs when
quantum degeneracy becomes important. Fluctuations are given by the
probability distribution of the macroscopic configurations with a
given atom number difference $n$. This distribution is binomial in the
classical gas regime, leading to $\langle n^2\rangle = N/4$. In the
degenerate regime, the entropy effect which favors small number
differences vanishes and, if each Fock state with a given $n$ is
equiprobable, $\langle n^2\rangle$ is as high as $N^2/12$.  The
crossover from super- to sub-binomial fluctuations comes from the
interaction energy cost associated with number fluctuations, which
eventually exceeds the available thermal energy.  This low temperature
regime is well described by the TMM, which predicts fluctuations
$\langle n^2\rangle =k_BT/E_C$, as a result of the equipartition
theorem. We heuristically extend the TMM to higher temperatures by
adding a binomial contribution for the thermal cloud, leading to
$\sqc^2=4k_BT/(NE_C)+1-N_{2m}/N$, where $N_{2m}$ is the condensate
fraction, {\it i.e.} the population of the two lowest modes, which has
to be determined independently. This formula implies the existence of
a maximum: on one hand the decrease of temperature increases $N_{2m}$,
which contributes to the super-binomial signal. On the other hand,
interactions that tend to lower the fluctuations dominate more and
more as the temperature decreases. Given that the maximum occurs at
some fraction of the critical temperature $T_c$, where a macroscopic
population in the two lowest mode appears, this maximum scales as
$k_BT_c/(NE_C)$, which is large for a cold gas in the weakly
interacting regime.

In order to account more accurately for the thermal cloud
contribution, we perform a multimode calculation in the classical
field approximation. The fluctuations are decomposed into a shot noise
term $N$ plus a term where the fields appear in the normal order. We
compute the latter by sampling the Glauber-P distribution that we
approximate by the classical distribution $P\propto \exp\{-\beta
E[\psi,\psi^*]\}$, where $E[\psi,\psi^*]$ is the Gross-Pitaevskii
energy functional~\cite{TheoYunAliceYvan}.  We include the first
quantum correction to the classical field, which has the effect to
change the shot noise term $N$ into $N-N_{2m}$. As expected, the
classical field approximation predicts an increase of fluctuations
starting above $T_c$, when the gas enters the degenerate regime. The
sub-binomial regime is out of reach of the classical field
approximation using the Glauber-P distribution. The results are shown
in Fig.~\ref{fig:theo}, and qualitatively reproduce the experimental
results (which were obtained in a dynamical process). The inset shows
$N_{2m}$ extracted from the classical field calculations, which we
also use in the TMM curve. In both models, fluctuations are largely
independent of the atom number $N$ for a given condensed fraction,
justifying the comparison with the experimental data.

These results show the interplay of quantum statistics and
interactions in a simple and fundamental finite-temperature system.
The double well can be seen as a ``nutshell'' version of the
Bose-Hubbard model whose more complex dynamics also lead to the Mott
insulator transition \cite{Greiner02a}. Our results also highlight the
generic features of spatial splitters for trapped BECs: interactions
allow for the creation of non-classical states of potential interest
for quantum metrology.  However, the same interactions will lead to
phase spreading with a rate proportional to $\sqc \sqrt{N} \, E_C$
after the splitting. Hence, tuning the interactions to reduce $E_C$
after splitting is a necessary step for the use of such a non-linear
splitter in a real interferometer.

 This work was funded in part by a EURYI grant. Our group is part of
 IFRAF. GEM acknowledges support from the Hertz Foundation.


\begin{thebibliography}{99}

\bibitem{Ketterle}
Y. Shin, M. Saba, T.~A. Pasquini, W. Ketterle, D.~E. Pritchard, and A.~E. Leanhardt,
Phys. Rev. Lett. {\bf 92}, 050405 (2004).
\bibitem{Anderson}
Y.-J. Wang, D.~Z. Anderson, V.~M. Bright, E.~A. Cornell, Q. Diot, T. Kishimoto,
  M. Prentiss, R.~A. Saravanan, S.~R. Segal, and S. Wu, Phys. Rev. Lett. {\bf
  94},  090405  (2005).
  \bibitem{Schumm}
  T. Schumm, S. Hofferberth, L.~M. Anderson, S. Wildermuth, S. Groth, I. Bar-Joseph,
J. Schmiedmayer, and P. Kr\"uger,
Nature Physics {\bf 1}, 57 (2005).

\bibitem{Oberthaler_Gati}
M. Albiez, R. Gati, J. F\"olling, S. Hunsmann, M. Cristiani, and M.~K.
  Oberthaler, Phys. Rev. Lett. {\bf 95},  010402  (2005).
\bibitem{Phase}
E.~M. Wright, D.~F. Walls, and J.~C. Garrison,
Phys. Rev. Lett. {\bf 77}, 2158 (1996);
J. Javanainen, M. Wilkens,
Phys. Rev. Lett. {\bf 78}, 4675 (1997);
M. Lewenstein, and L. You, Phys. Rev. Lett. {\bf 77}, 3489 (1996);
Y. Castin and J. Dalibard, Phys. Rev. A \textbf{55}, 4330 (1997).

\bibitem{Squeezing}
A. J. Leggett and F. Sols,
Phys. Rev. Lett. {\bf 81}, 1344 (1998).
\bibitem{Kitagawa}
M. Kitagawa and M. Ueda, Phys. Rev. A {\bf 47},  5138  (1993).
\bibitem{Squeezing_Ketterle}
G.-B. Jo, Y. Shin, S. Will, T.~A. Pasquini, M. Saba, W. Ketterle, D.~E.
  Pritchard, M. Vengalattore, and M. Prentiss, Phys. Rev. Lett. {\bf 98},
  030407  (2007).
  \bibitem{BPhillipps}
J. Sebby-Strabley, B.~L. Brown, M. Anderlini, P.~J. Lee, W.~D. Phillips, J.~V. Porto, and P.~R. Johnson, Phys. Rev. Lett. {\bf
  98}, 200405 (2007).
\bibitem{Squeezing_MKO}
J. Est{\`e}ve, C. Gross, A. Weller, S. Giovanazzi, and M. Oberthaler, Nature
  {\bf 455},  1216  (2008).

\bibitem{TMM_Ananikian}
D. Ananikian and T. Bergeman, Phys. Rev. A {\bf 73},  013604  (2006) and references therein.

\bibitem{DGO}
G. Reinaudi, T. Lahaye, Z. Wang, and D. Gu{\'e}ry-Odelin, Optics Letters {\bf
  32},  3143  (2007).

\bibitem{Davis}
A.~J. Ferris, M.~J. Davis, arXiv:1001.2041.

\bibitem{Bodet}
C. Bodet, J. Est{\`e}ve, M.~K. Oberthaler, and T. Gasenzer, arXiv:1002.2504

\bibitem{TheoYunAliceYvan}
A. Sinatra, Y. Castin, and Yun Li, arXiv:1003.0761, to appear in
Phys. Rev. A.

\bibitem{EsteveBouchoule_bunching}
J. Est{\`e}ve, J.-B. Trebbia, T. Schumm, A. Aspect, C.~I. Westbrook, and
  I. Bouchoule, Phys. Rev. Lett., {\bf 96}, 130403 (2006).

\bibitem{Greiner02a} M. Greiner, O. Mandel, T.~W. H{\"a}nsch, and
  I. Bloch, Nature {\bf 415}, 39 (2002).
\end{thebibliography}

\end{document}